\documentclass[universe,review,accept,oneauthors,pdftex]{Definitions/mdpi} 
\usepackage{color}

\newcommand\sersic{S\'ersic}
\newcommand\propol{{\em propol}} 

\firstpage{1} 
\pubvolume{1}
\issuenum{1}
\articlenumber{0}
\pubyear{2022}
\copyrightyear{2022}
\datereceived{\today} 
\dateaccepted{} 
\datepublished{} 
\hreflink{https://doi.org/} 

\Title{The principle of maximum entropy and the distribution of mass in galaxies}

\TitleCitation{Maximum entropy and mass distribution of galaxies}

\Author{Jorge S\'anchez Almeida$^{1,2}$*\orcidA{0000-0003-1123-6003}}

\AuthorCitation{S\'anchez Almeida, J.}

\address{%
$^{1}$ \quad  Instituto de Astrof\'\i sica de Canarias, La Laguna, Tenerife, E-38200, Spain\\
$^{2}$ \quad Departamento de Astrof\'\i sica, Universidad de La Laguna, Tenerife, Spain}

\corres{Correspondence: jos@iac.es)}

\abstract{
 We do not have a final answer to the question of  {\em  why galaxies choose a particular internal mass distribution.} Here we examine whether the distribution is set by thermodynamic equilibrium (TE). Traditionally,  TE is discarded for a number of reasons including the inefficiency of two-body collisions to thermalize the mass distribution in a Hubble time, and the fact that the mass distribution maximizing the classical Boltzmann-Gibbs entropy is unphysical. These arguments are questionable. In particular, when the Tsallis entropy that describes self-gravitating systems is used to define TE, the mass distributions that result (i.e., the polytropes) are physically sensible. This work spells out this and other arguments for TE, and presents the polytropes and their properties. It puts forward empirical evidence for the mass distribution observed in galaxies to be consistent with polytropes. It compares polytropes with \sersic\ functions and it shows how the DM halos resulting from cosmological numerical simulations become polytropes when efficient collisions are allowed. It also discusses pathways to thermalization bypassing two-body collisions. It finally outlines future developments including deciphering whether or not DM particles collide efficiently. 
}

\begin{document}
%

 %
%
  %
  %
%

\section{Purpose and organization of this review}\label{sec:intro}
Galaxies are self-gravitating structures approaching mechanical equilibrium, where kinetic energy ($\langle T\rangle$) and gravitational energy ($\langle U\rangle$) tend to balance each other to meet the virial condition ($2\langle T\rangle+\langle U\rangle=0$). The constraint imposed by this condition is very loose, though. It allows for a large number of ways to distribute the mass internally \citep[][]{2008gady.book.....B}. This freedom contrasts with the fact that real galaxies choose only very specific internal mass distributions, namely, those consistent with stellar mass surface density profiles looking like S\'ersic functions \citep[e.g.,][]{1968adga.book.....S,1993MNRAS.265.1013C,2001MNRAS.326..869T,2003ApJ...594..186B,2005PASA...22..118G,2012ApJS..203...24V}. These functions embrace from exponential disks  \cite[observed in dwarf galaxies; e.g.,][]{1994A&AS..106..451D} to  {\em de Vaucouleurs} $R^{1/4}$-profiles \cite[characteristic of massive ellipticals; e.g.,][]{1948AnAp...11..247D}.

{\em  Why do galaxies choose this particular mass distribution?} Although it sounds a trivial question, we do not know the answer yet. It could be due either to satisfy some fundamental physical principle (e.g.,  thermodynamic equilibrium; hereinafter TE) or to the very particular initial conditions that gave rise to the system \citep{2008gady.book.....B}. Contrarily to what one may naively think, the mass distribution in galaxies is commonly explained as the outcome of cosmological initial conditions \citep[][and  Sect.~\ref{sec:nfwproperties}]{2004MNRAS.352.1109A,2004MNRAS.355.1217H,2014ApJ...790L..24C,2015ApJ...805L..16N,2017MNRAS.465L..84L,2020MNRAS.495.4994B}. The option of a  fundamental process like TE determining the internal configuration is traditionally disregarded with two different arguments which, however, seem to become frailer as time goes by and more observational and theoretical work is available. 

The first argument against TE has to do with the nature of dark matter (DM). According to the current concordance cosmological model,  DM provided most of the gravity that allowed galaxies to form in time, and now it holds them together. In the simplest model, DM is collision-less and cold (CDM). Under these hypotheses, the timescale for the DM particles to reach TE is set by the two-body relaxation timescale, which is extremely long; typically much larger than the Hubble time \citep[i.e., than the age of the Universe;][]{2008gady.book.....B,2003MNRAS.338...14P,2019MNRAS.488.3663L,2021MNRAS.504.2832S}. Thus, taken literally, this result would discard mass distributions arising from TE.  However, the collision-less nature assumed for the DM causes some of the so-called small-scale problems of the CDM model \citep[e.g.,][]{2015PNAS..11212249W,2017ARA&A..55..343B,2017Galax...5...17D}, in particular, the {\em cusp--core problem}. Simulated CDM haloes have {\em cusps} in their central mass distribution \citep[e.g.,][; see Sect.~\ref{sec:nfwproperties}]{1997ApJ...490..493N,2020Natur.585...39W} which disagree with the central plateau or {\em core} often observed in galaxies \citep[e.g.,][]{2015AJ....149..180O,2021arXiv210503435C,2021ApJ...921..125S}. This well known difficulty of the CDM model is bypassed invoking  physical processes which essentially shorten the timescale to reach TE, making it  shorter than the Hubble time.  The proposed pathways to thermalization are of very different nature (Sect~\ref{sec:thermalization}):  feedback of the baryons on the DM particles through gravitational forces \citep[][]{2010Natur.463..203G,2014MNRAS.437..415D,2020MNRAS.499.2912F},  scattering with massive gas clumps \citep[][]{2013ApJ...775L..35E,2019MNRAS.489.5919S}, forcing by a central bar \citep{1971ApJ...168..343H}, merger of two super massive black holes \citep[SMBHs;][]{2006AJ....132.2685M,2021MNRAS.502.4794N,2021ApJ...921..125S}, or assuming an artificially large DM to DM collision cross section \citep{2000PhRvL..84.3760S,2001ApJ...547..574D,2015MNRAS.453...29E}.
Whether or not it is collision-less, the DM seems to find a way for thermalization faster than the inefficient two-body relaxation collisions, an observational fact that weakens the original criticism.  

The second argument against TE setting the internal structure of galaxies is related to the use of the  classical Boltzmann-Gibbs entropy to define equilibrium. Following the principles of statistical physics, the structure in TE corresponds to the most probable configuration of a self-gravitating system and, thus, it should result from maximizing its entropy\footnote{This is precisely the meaning given in the paper to the term {\em thermodynamic equilibrium}. It is used for  distributions that maximize the funcional describing the entropy of the system.}. The use of the classical Boltzmann-Gibbs entropy to characterize  self-gravitating systems leads to a distribution with infinite mass and energy \citep[][and Sect.~\ref{sec:maxentro}]{2008gady.book.....B,2008arXiv0812.2610P}, disfavoring TE as explanation. However, this difficulty of the theory has been overcome as follows. In the standard Boltzmann-Gibbs approach, the long-range gravitational forces that govern self-gravitating systems are not taken into account. These forces are not subordinate but a fundamental ingredient of the physical system. Fortunately, systems with long-range interactions admit long-lasting meta-stable states described by the Tsallis ($S_q$) non-additive entropy \citep[][]{1988JSP....52..479T,2009insm.book.....T,2005PhyA..356..419C}.  In particular, the maximization under suitable constraints of the Tsallis entropy of a self-gravitating N-body system leads to a polytropic distribution \citep{1993PhLA..174..384P,2005PhyA..350..303L}, which can have finite mass and a shape resembling the DM distribution found in numerical simulations of galaxy formation \citep[][and Sect.~\ref{sec:nfwproperties}]{2004MNRAS.349.1039N,2009PhyA..388.2321C,2013MNRAS.428.2805A,2021MNRAS.504.2832S,2005ApJ...624L..85M}. The polytropic shape has lately gained practical importance because of its association with real self-gravitating astrophysical objects (Sect.~\ref{sec:obs_support}).  The mass density profiles in the centers of dwarf galaxies are reproduced by polytropes without any tuning or fitting \citep{2020A&A...642L..14S}. Polytropes  also explain the stellar surface density profiles observed in galaxies \citep{2021ApJ...921..125S} and in globular clusters  \citep{trujillo21}. 

In short, the arguments against TE setting galaxy shapes are questionable. Moreover, empirical evidence supports that the mass distribution expected from this equilibrium is indeed observed in self-gravitating astronomical objects. These two facts justify the writing of the present review. It basically expands the above arguments, gathering the works suggesting the fundamental role played by TE and spelling out the arguments leading to this conclusion. 

The paper is organized as follows:
Sect.~\ref{sec:maxentro} explains in detail the density profiles to be expected in self-gravitating systems of maximum entropy. Firstly, using the Boltzmann-Gibbs entropy (Sect.~\ref{sec:BGentropy}) and later on (Sect.~\ref{sec:tsallis}) the Tsallis entropy that yields polytropic density profiles. The physically relevant properties of polytropes are summarized in Sect.~\ref{properties_pol}. In order to compare polytropes with observations, they have to be projected in the plane-of-the-sky, an exercise carried out in Sect.~\ref{sec:propols}.
Section~\ref{sec:sersic} discuses the relation between projected polytropes (\propol s) and \sersic\ profiles.
Section~\ref{sec:nfwproperties} briefly mentions the relation between polytropes and the profiles inferred from N-body numerical simulations of structure formation in the Universe. 
Section~\ref{sec:obs_support} summarizes all the evidence for polytropic profiles reproducing real galaxies, thus providing observational support for the theory described in previous sections.
Section~\ref{sec:thermalization} discusses the different pathways that have been put forward to explain how galaxies or parts of galaxies may have reached TE already.
Finally, Sect.~\ref{sec:conclusions} summarizes the main results presented here and lists several challenges to be addressed in the next years.

%
%
\section{Self-gravitating systems of maximum entropy}\label{sec:maxentro}

This section follows arguments and notation taken from \citep[][]{2008gady.book.....B,1993PhLA..174..384P,2021ApJ...921..125S}.

\subsection{Solution using the classical Boltzmann-Gibbs entropy: isothermal sphere}\label{sec:BGentropy}

Assume a spherically symmetric self-gravitating system of identical particles. The classical Boltzmann-Gibbs entropy of the system is defined as
\begin{equation}
  S = -\int f\ln f\,d^6{\bf w}+C,
  \label{eq:bgentopy}
\end{equation}
with $f$ the distribution function (DF), $d^3{\bf w}$ the volume element in the 6D phase space (of position and velocity), and $C$ an integration constant. The integral extends to all the phase space. The maximum entropy solution would be the  distribution $f$ that maximizes $S$ under the constraints that the total  mass $M$ and energy $E$ are constant,
\begin{equation}
  M = \int f\,  d^6{\bf w},
  \end{equation}
\begin{equation}
 E = \int \epsilon\,f\,  d^6{\bf w},
\end{equation}
  with $\epsilon = \Phi+v^2/2$ the total energy per unit mass. The symbols $v$ and $\Phi$ stand for the velocity and the gravitational potential, respectively.  Using Lagrange multipliers and variational calculus, the condition for $S$ to be extreme follows from,
  \begin{equation}
    \frac{\partial}{\partial f} [f\ln f + \alpha f+ \beta \epsilon f] = 0,
    \label{eq:euler-lagrange}
\end{equation}
with $\alpha$  and $\beta$ two Lagrange multipliers. Satisfying this condition requires $f$ to be an exponential function of the energy, namely,
\begin{equation}
  \ln f =-(1+\alpha+\beta\epsilon).
  \label{eq:exp}
\end{equation}
Since the system is self-gravitating, the gravitational potential, $\Phi$, and the density, 
\begin{equation}
  \rho=\int f\, d^3{\bf v},
  \label{eq:def_rho}
\end{equation}
corresponding to $f$ have to satisfy the Poisson equation for spherically symmetric systems,
\begin{equation}
  \nabla^2 \Phi = \frac{1}{r^2}\frac{d}{dr}\Big(r^2\frac{d\Phi}{dr}\Big) =4\pi G\rho,
  \label{eq:poisson}
\end{equation}
with $G$ the gravitational constant and $r$ the radial distance to the center of the potential. Equations ~(\ref{eq:exp}), (\ref{eq:def_rho}), and (\ref{eq:poisson}) are better solved in terms of the relative potential $\Psi = \Phi_0-\Phi$ and the relative energy $\varepsilon =\Phi_0-\epsilon$, with the constant $\Phi_0$ chosen so that $\varepsilon > 0$ for $f> 0$ \citep[see ][Chap.~4]{2008gady.book.....B}. Then,
\begin{equation}
\ln f = \beta(\Psi-\frac{v^2}{2})+C',
\end{equation}
and so through the relation~(\ref{eq:def_rho}),
\begin{equation}
\ln\rho = \beta \Psi+C'',
\end{equation}
which transforms Eq.~(\ref{eq:poisson}) into
\begin{equation}
  \frac{d}{dr}\Big(r^2\frac{d\ln\rho}{dr}\Big)=- 4\pi G\beta\, r^2\rho.
  \label{eq:iso_sphere}
\end{equation}
The previous equation describes an {\em isothermal sphere}, i.e., it is formally identical to the density structure arising in a self-gravitating ideal gas of constant temperature $T$, with  
\begin{equation}
\beta = \frac{m_g}{k T},
\end{equation}
where $m_g$ is the mass of each gas molecule and $k$ the Boltzmann constant. In general, \mbox{Eq.~(\ref{eq:iso_sphere})} has to be integrated numerically, however,  it admits a solution called {\em singular isothermal sphere},
\begin{equation}
  \rho\propto \frac{1}{r^2},
    \label{eq:iso_sphere_sing}
\end{equation}
which is important in the present context since it describes the asymptotic behavior of all solutions of Eq.~(\ref{eq:iso_sphere}) when $r\rightarrow\infty$ \citep{1967aits.book.....C,2008gady.book.....B}. This implies that the total mass of the system is always infinity,
\begin{equation}
  M\propto \lim_{r\rightarrow\infty} \int_0^r r'^2\rho(r')dr' \simeq \lim_{r\rightarrow\infty} \int_0^rdr' \rightarrow\infty.   
  \end{equation}
  If rather than the total mass one works out the total energy, it also turns out to be infinity. Finally, the entropy of the system (Eq.~[\ref{eq:bgentopy}]) tends to infinity as well,
  \begin{displaymath}
    S \propto -\lim_{r\rightarrow\infty}  \int_0^r \int_0^\infty (\ln\rho-\beta\frac{v^2}{2})\, \rho\, \exp(-\beta\frac{v^2}{2})\,v^2dv\, r'^2dr'
 \end{displaymath} 
 \begin{equation}
  \simeq  \int_0^\infty \Big[\lim_{r\rightarrow\infty}  \int_0^r(2\ln r'+\beta\frac{v^2}{2}) dr'\Big]\,\exp(-\beta\frac{v^2}{2})\,v^2dv\, \rightarrow\infty.
\end{equation}
Therefore, any self-gravitating system asked to reach maximum Boltzmann-Gibbs entropy has infinite mass, infinite energy, and infinite entropy, and so, it is non-physical. \citet[][their Sect.~4.10.1]{2008gady.book.....B} explain that no $f$ with finite $M$ and $E$ maximizes $S$ because a trivial re-arrangement of the mass distribution can augment $S$ boundlessly.  Given $M$ and $E$, $S$ always can be increased by increasing the degree of central concentration and then transferring the resulting gain of potential energy to an arbitrarily small amount of mas placed in a large outer envelope.

%
%
\begin{figure}
\centering
\includegraphics[width=0.65\linewidth]{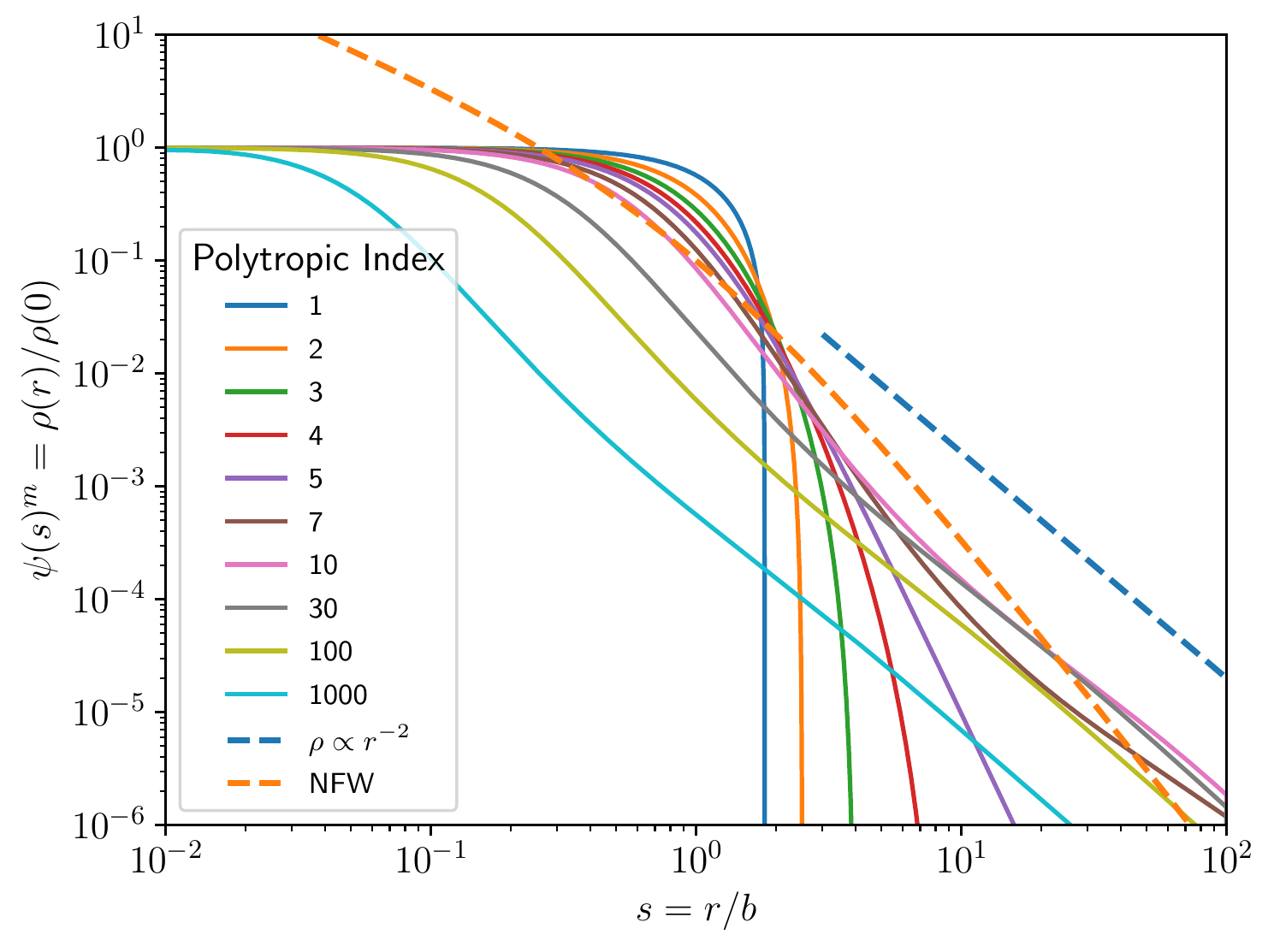}
\caption{
Polytropes resulting from solving the Lane-Emden Eq.~(\ref{eq:lane_emden}).  These curves just provide the shape of the mass density profile, which can be shifted horizontally and vertically by arbitrary amounts (set by the constants $b$ and $\rho(0)$, respectively). The color code of the solid lines gives the polytropic index $m$ as indicated in the inset. The curve with $m=1000$ approximately corresponds to the shape of an isothermal sphere. For reference, the figure  includes  $\rho\propto r^{-2}$ (the blue dashed line), which represents the asymptotic behavior of high-$m$ polytropes, and also shows a NFW profile (the orange dashed line) \citep[][]{1997ApJ...490..493N}. 
}
\label{fig:lane_emden}
\end{figure}
%
\begin{figure}
\centering 
\includegraphics[width=0.65\linewidth]{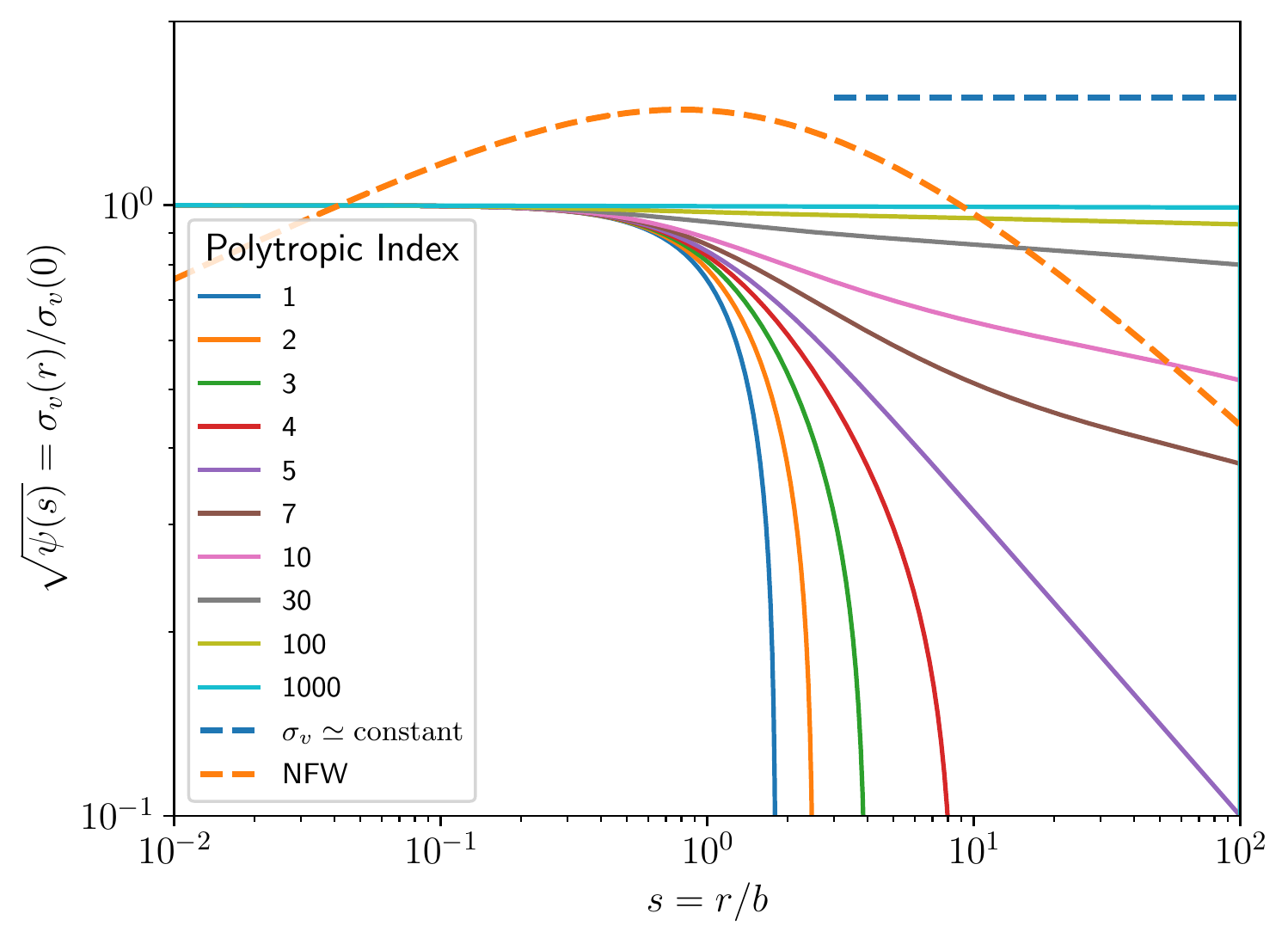}
\caption{
Velocity dispersion corresponding to the  polytropic radial density variation represented in Fig.~\ref{fig:lane_emden}. These curves just provide the shape of the profile, which can be shifted horizontally and vertically by arbitrary amounts (set by the constants $b$ and $\sigma_v(0)$, respectively). The color code  is the same as that used in Fig.~\ref{fig:lane_emden}. The curve with $m=1000$ approximately models the shape of an isothermal sphere.  For reference, the plot includes $\sigma_v \simeq {\rm constant}$ (the blue dashed line), which represents the asymptotic behavior of high-$m$ polytropes,  and also shows the velocity dispersion of the NFW profile in Fig.~\ref{fig:lane_emden} (as worked out in \citep{2004ApJ...602..162H}). Note the huge difference in dynamical range between the radial density variations ($\sim 10^6$; Fig.~\ref{fig:lane_emden}) and the corresponding velocity dispersion variations shown in this figure ($\sim 10$).
}
\label{fig:lane_emden_disp}
\end{figure}
%
%
\subsection{Solution using the Tsallis entropy: polytropes}\label{sec:tsallis}
The Boltzmann-Gibbs entropy neglects long range interactions leading to the inconsistencies pointed out in Sect.~\ref{sec:BGentropy}.  Back in \citeyear{1988JSP....52..479T}, \citet{1988JSP....52..479T} introduced another entropy, now called Tsallis entropy,
\begin{equation}
  S_q = - \int f\,\frac{1-f^{q-1}}{q-1}\, d^6{\bf w}+C'''.
\end{equation}
The Tsallis entropy has been successfully used in many different physical contexts going from plasmas and particle physics to geophysics and astrophysics \citep[e.g., ][Sect.~6]{2011Entrp..13.1765T}. It works because $S_q$ is able to describe systems having long range interactions among their constituents \citep{2009insm.book.....T,2011Entrp..13.1765T}. Tsallis entropy actually represents a full family of entropies since it depends on the numerical coefficient $q$. In the special case when $q\rightarrow 1$ it becomes the Boltzmann-Gibbs entropy in Eq.~(\ref{eq:bgentopy})\footnote{Consider the identity $(1-f^{q-1})/(q-1) = (1-\exp[(q-1)\ln f])/(q-1)$, expand $\exp$ in its Taylor series, and take the limit $q\rightarrow 1$.}.
Following the standard approach described in Sect.~\ref{sec:BGentropy},  \citet{1993PhLA..174..384P} worked out the DF $f$  that minimizes $S_q$ for a system with constant mass and energy. In this case the Lagrange Eq.~(\ref{eq:euler-lagrange}) becomes, 
  \begin{equation}
    \frac{\partial}{\partial f} [ f\,\frac{1-f^{q-1}}{q-1} + \alpha f+ \beta \epsilon f] = 0,
    \label{eq:needed}
\end{equation}
which provides,
\begin{equation}
  f = \Big[\frac{1}{q}\Big( 1- (\alpha+\beta\epsilon)(1-q)\Big)\Big]^{\frac{1}{q-1}},
  \end{equation}
  or, redefining $\Phi_0$,
  \begin{equation}
    f \propto \Big[\Phi_0-\epsilon\Big]^{\frac{1}{q-1}}=\Big[\Psi-\frac{v^2}{2}\Big]^{\frac{1}{q-1}}.
    \label{eq:alsoneeded}
  \end{equation}
The DF in Eq.~(\ref{eq:alsoneeded}) corresponds to a {\em polytrope} \citep[Eqs.~(4.41) and  (4.83) in ][]{2008gady.book.....B} with the polytropic index $m$ set by $q$ as
  \begin{equation}
    m=\frac{3}{2} + \frac{1}{q-1},
    \label{eq:equivalence}
  \end{equation}
and with the density (Eq.~[\ref{eq:def_rho}]) given by
 \begin{equation}
\rho \propto \Psi^m.
 \end{equation}
 For the system to be self-gravitating, the density $\rho$ and the gravitational potential $\Phi$ ($=\Psi-\Phi_0$)  have to be linked through the Poisson Eq.~(\ref{eq:poisson}). After some algebra, the Poisson equation can be re-written as the so-called  Lane-Emden equation \citep{1967aits.book.....C,2008gady.book.....B}, which in practice is taken to define polytropes. Explicitly, a polytrope  of index $m$ is defined as the spherically-symmetric self-gravitating structure resulting from the solution of the Lane-Emden equation for the (normalized) gravitational potential, namely,
\begin{equation}
  \frac{1}{s^2}\frac{d}{ds}\Big(s^2\frac{d\psi}{ds}\Big)=
  \begin{cases}
    -3\psi^m& \psi > 0,\\
    0 & \psi \le  0.
  \end{cases}
  \label{eq:lane_emden}
 \end{equation}
 The symbol  $\psi$ stands for the normalized relative potential ($\propto \Psi $), $s$ represents the scaled radial distance,
\begin{equation}
  r = b\, s,
   \label{eq:radius}
 \end{equation}
 and the density at radial distance $r$ is recovered from $\psi$ as
 \begin{equation}
   \rho(r) = \rho(0)\,\psi(s)^m,
   \label{eq:densityle}
 \end{equation}
 where $\rho(0)$ and $b$ are two constants. In general, Eq.~(\ref{eq:lane_emden}) has to be solved numerically under the initial conditions
\begin{equation}
\psi(0)=1 ~~~{\rm and} ~~~ d\psi(0)/ds =0. \label{eq:ini_cond}
\end{equation}
Solutions with $d\psi(0)/ds \not= 0$ are discarded because they have infinite central density and total mass \citep[e.g.,][]{2008gady.book.....B}\footnote{Plugging into Eq.~(\ref{eq:lane_emden}) the expansion of $\psi$ around $s=0$, $\psi(s)\simeq \psi(0)+[d\psi(0)/ds]\,s+\dots$, one ends up with the identity $\psi(0)\simeq -(3/2)[d\psi(0)/ds]\,s^{-1}$, which implies that either $d\psi(0)/ds=0$ or $\psi(0)\rightarrow\infty$ when $s\rightarrow\infty$.}. 

In contrast with the maximum entropy solutions for the Boltzmann-Gibbs entropy, the maximum entropy solution for the Tsallis entropy are physically meaningful provided $m$ (and so $q$) is within a fairly narrow range of values,
 \begin{equation}
  3/2 < m \leq 5.
  \label{eq:nlimits}
\end{equation}
The upper limit ($m\leq 5$) comes from requesting the mass to be finite \citep{1967aits.book.....C,2008gady.book.....B}. The cause for the lower limit is more subtle. Polytropes with $m < 3/2$ have an $f$ which increases with increasing energy per particle, which is physically unreliable  \citep{1993PhLA..174..384P}.
In addition, the stability of polytropes is analyzed in various works \citep{2002A&A...386..732C,2002PhyA..307..185T,2008gady.book.....B} which set the condition $m> 3/2$ for the polytropes to be stable for radial pulsations.

Figure~\ref{fig:lane_emden} shows the polytropes resulting from solving the Lane-Emden Eq.~(\ref{eq:lane_emden}).  These curves just provide the shape of the density profile, which can be shifted horizontally and vertically by arbitrary amounts (Eqs.~[\ref{eq:radius}] and [\ref{eq:densityle}]). The color code of the solid lines gives the polytropic index as indicated in the inset. The curve with $m=1000$ approximately corresponds to the shape of an isothermal sphere, which is the limiting solution for $m\rightarrow\infty$.

The DF $f$ in Eq.~(\ref{eq:def_rho}), integrated over all velocities at a given point, renders the density in Eq.~(\ref{eq:densityle}). The same exercise can be used to derive the velocity dispersion $\sigma_v$,
\begin{equation}
  \rho\,\sigma_v^2=\int  v^2\,f \,d^3{\bf v},
\end{equation}
and it leads to
\begin{equation}
  \sigma_v^2(r) = \sigma_{v}^2(0)\, \psi,
  \label{eq:poly_disp}
 \end{equation}
with the velocity dispersion in the center of the gravitational potential, $\sigma_{v}^2(0)$, connected with the central density through 
 \begin{equation}
  \sigma^2_{v}(0) = 4\pi G\rho(0) b^2/(m+1).
\end{equation}
Since the physical systems is axi-symmetric, the velocity dispersion in one particular direction (e.g.,$z$) is just $\sigma_{vz}  = \sigma_v/\sqrt{3}$.

Figure~\ref{fig:lane_emden_disp} shows the radial variation of the velocity dispersion corresponding to the polytropic radial density variation represented in Fig.~\ref{fig:lane_emden}. For reference, the figure also includes a curve representing the asymptotic behavior of high-$m$ polytropes (i.e., $\sigma_v\simeq$~constant),  and the velocity dispersion corresponding to a NFW profile. Polytropes and NFW profiles also behave very differently in terms of their velocity dispersion \citep{2007ApJ...655..847B}.   
%

%
\subsubsection{Properties of the polytropes}\label{properties_pol}

The solutions of the Lane-Emden equation ($\psi$ in Eq.~[\ref{eq:lane_emden}]) are well studied in literature because of their connection with the stellar structure produced by polytropic gases \citep[][]{1964ApJS....9..201F,1967aits.book.....C,2004ASSL..306.....H}. Some of these properties are of relevance for the study of the structure of galaxies, and they are compiled next. 

\smallskip\noindent$\bullet$ {\em Analytic solutions.}
The case $m=5$ is known as the \citeauthor{schuter84} sphere or \citeauthor{1911MNRAS..71..460P} model and has an analytic solution,
\begin{equation}
  \psi(s) = \frac{1}{\sqrt{1+s^2}},
  \label{eq:poly5}
  \end{equation}
  which extends to infinity but has finite mass. Analytic solutions also exists for $m=0$ and $m=1$ \citep[][]{1967aits.book.....C}. They are useful for the purpose of testing numerical solutions of Eq.~(\ref{eq:lane_emden}), but they are no so relevant in the present context because of Eq.~(\ref{eq:nlimits}).

  \smallskip\noindent$\bullet$ {\em Connection with the isothermal sphere.} Since the Boltzmann-Gibbs entropy approaches the Tsallis entropy in the limit $q\rightarrow 1$, and this limit corresponds to $m\rightarrow\infty$ (see Eq.~[\ref{eq:equivalence}]), one can recover the isothermal sphere (Eqs.~[\ref{eq:iso_sphere}] and [\ref{eq:iso_sphere_sing}]) taking the limit of the Lane-Emden Eq.~(\ref{eq:lane_emden}) when  $m\rightarrow\infty$ \citep{2001MNRAS.328..839H}.

  \smallskip\noindent$\bullet$ {\em Analytic approximations}. Most polytropic indexes do not admit an analytic solution and either Eq.~(\ref{eq:lane_emden}) has to be integrated numerically or one has to resort to one of the analytic approximations existing in literature. The original approximations date back to pre-computer days when polytropes were fundamental to model stelar structure \citep{1967aits.book.....C}. However, these approximations may still be useful nowadays to fit large datasets of galaxies. Fitting requires to evaluate polytropes thousands of times, and this computation usually determines the speed of the algorithm.
\citet{1967aits.book.....C} expands $\psi$ as a polynomial around $s=0$. The first terms are
\begin{equation}
  \psi(s) \simeq 1-\frac{1}{2}\,s^2+\frac{3\,m}{40}\,s^4-\dots,\label{eq:expansion} 
\end{equation}
an expression which holds for $s\lesssim 1$. Through Eq.~(\ref{eq:densityle}), one recovers the volume density.
An expansion reaching out to the term $s^{10}$ is explicitly given by \citet[][Eq. (2.4.24)]{2004ASSL..306.....H}.  This approximation eventually breaks down for $s$ large enough and so other alternatives have being worked out. Approximations in terms of exponentials have being tried \citep[][]{1964ApJS....9..201F,1987Ap&SS.132..393B}. Through an Euler transformation of Eq.~(\ref{eq:lane_emden}), \citet{2001MNRAS.328..839H}  works out a series that converges all the way to the outer radius. A comprehensive review of the possibilities put forward in literature are compiled by \citet{2004ASSL..306.....H}.  Among them, it is shown that for large $s$, $\psi(s)$  approaches  the solution
\begin{equation}
  \psi(s) \simeq A_m\,s^{2/(1-m)}
  \label{eq:larges}
\end{equation}
in an oscillatory manner, with  the coefficient $A_m$ depending just on $m$ \citep[see][Eq.~(2.4.88)]{2004ASSL..306.....H}. Thus, due to Eqs.~(\ref{eq:densityle}), (\ref{eq:radius}), and (\ref{eq:larges}),
\begin{equation}
\rho\propto r^{-2}
\end{equation}
for $m \gg 1$.  Figure~\ref{fig:lane_emden} also includes for reference the curve $\rho\propto r^{-2}$ (the blue dashed line) to represent the asymptotic behavior of high-$m$ polytropes. Likewise, the asymptotic behavior of the velocity dispersion for $m \gg 1$ inferred from Eqs.~(\ref{eq:poly_disp}) and (\ref{eq:larges}) is a constant independent of $s$, as it should be in this limit when polytropes become isothermal spheres.    

%
\smallskip\noindent$\bullet$ {\em Central cores}. All politropes have a {\em core} (a central plateau in the density distribution) with the same shape independently of the index $m$ \citep{2020A&A...642L..14S,2021MNRAS.504.2832S}. Equations~(\ref{eq:densityle}) and (\ref{eq:radius}), and the first two terms in the approximation~(\ref{eq:expansion}) lead to, 
 \begin{equation}
   \frac{\rho(r)}{\rho_\alpha} \simeq
   \Big[1+\frac{\alpha}{2m}\Big(1-\frac{r^2}{r_\alpha^2}\Big)\Big]^m
   \simeq 1+\frac{\alpha}{2}\Big(1-\frac{r^2}{r_\alpha^2}\Big),
   \label{eq:central2}
 \end{equation}
 with the characteristic radius, $r_\alpha$, and characteristic density,  $\rho_\alpha=\rho(r_\alpha)$, defined in terms of a particular value ($-\alpha$) for the logarithmic derivative of the density profile,
 \begin{equation}
   \frac{d\ln\rho}{d\ln r}(r_\alpha)=-\alpha.
   \label{eq:logder}
 \end{equation}
 The second approximate identity in the right-hand side of Eq.~(\ref{eq:central2}), which holds for $\alpha \ll  2m$, indicates that after normalization  by $\rho_\alpha$ and $r_\alpha$, all polytropes collapse to a single shape independent of $m$. In other words, except for a trivial normalization, all the polytropes look the same in their cores. This property is followed by haloes from numerical simulations of self-gravitating systems (Sect.~\ref{sec:nfwproperties}) as well as in real galaxies (Sect.~\ref{sec:obs_support}).

%
\smallskip\noindent$\bullet$ {\em Finite size.}
We also note that for $m < 5$ there is always a truncation, i.e., $\rho$ goes to zero at a finite radius \citep{1967aits.book.....C}. Thus, these polytropes have a finite size. The radii have to be computed numerically, and are tabulated elsewhere \citep[e.g.,][Table~4]{1967aits.book.....C}.

%
\smallskip\noindent$\bullet${\em Relation with polytropic gases.} As we explain above, the properties of polytropes are well known because of the formal equivalence with polytropic gases, traditionally used to model stars analytically \cite[][]{1967aits.book.....C}. They are formally identical provided that the exponent $\gamma$ that defines the relation between pressure ($P$) and $\rho$,
\begin{equation}
P\propto \rho^\gamma,
\end{equation}
is related with the polytropic index in the Lane-Emden Eq.~(\ref{eq:lane_emden}) as \citep[e.g.,][]{2008gady.book.....B},
\begin{equation}
\gamma = 1+\frac{1}{m}.
\end{equation}
In this equivalence, $\sigma_v^2$ plays the role of gas temperature ($T$) so that the perfect gas law, $P\propto \rho\,T$, turns into,
\begin{equation}
  P\propto \rho\,\sigma_v^2.
  \label{eq:equiv_temp}
\end{equation}

%
\smallskip\noindent$\bullet${\em Relation between density and velocity dispersion.}%
Putting together Eqs.~(\ref{eq:densityle}) and (\ref{eq:poly_disp}), one finds a one-to-one relation between $\rho$ and $\sigma_v$, namely,
\begin{equation}
  \rho(r) = \rho(0)\,[\sigma_v(r)/\sigma_{v}(0)]^{2m}.
  \label{eq:contrast}
\end{equation}
%
%
\subsubsection{Plane-of-the-sky projected polytropes}\label{sec:propols}
\begin{figure}[h!]
\centering 
\includegraphics[width=0.65\linewidth]{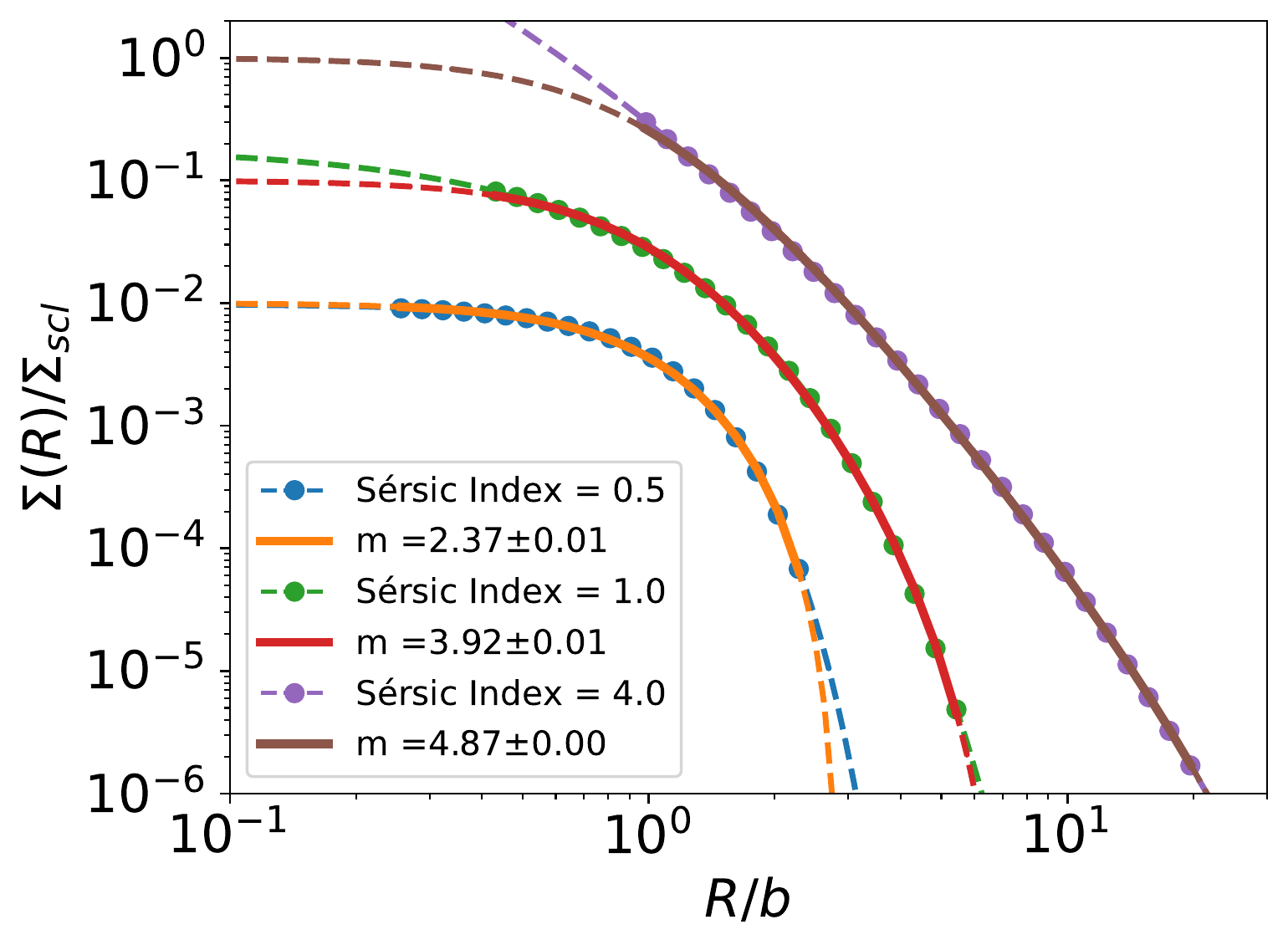}
\caption{Representative \sersic\ profiles (symbols and dashed lines of the same color) fitted with \propol s (solid and dashed lines of the same color).  The symbols mark the range of radii included in the fit. Except for \sersic\ index 0.5, cores have been excluded since the cores of \sersic\ profiles  and \propol s are different. Outside the core, the agreement is well within any realistic observational error, an agreement holding for up to a factor $\sim\,20$ in radius and $\sim\,10^5$ in surface density.  The original \sersic\ index and the corresponding polytropic index are given in the inset. For display purposes, all profiles have different normalization ($\Sigma_{scl}$) and are scaled radially to the scale-length of the corresponding polytrope ($b$).}
\label{fig:lane_emden_fitsersic}
\end{figure}

The polytropes describe a 3D mass distribution whereas the comparison with astronomical observations is often made in terms of the surface density projected in the plane of the sky  (see Sects.~\ref{sec:sersic} and \ref{sec:obs_support}). Thus, the 2D projection of the polytropes \citep[called \propol s by][]{2021ApJ...921..125S} deserves analysis. The surface density $\Sigma(R)$ corresponding to the volume density $\rho(r)$ is given by its Abel transform,  with $R$ the projected distance from the center \citep[e.g.,][]{2008gady.book.....B}. Then, $\Sigma(R)$ can be expressed in terms of the normalized Abel transform $f(x,m)$, 
\begin{equation}
  \Sigma(R) = a\,f(R/b,m),
  \label{eq:needlabel}
\end{equation}
with
\begin{equation}
  f(x,m) = 2\,\int_x^\infty \frac{s\,\psi^m(s)\,ds}{\sqrt{s^2-x^2}},
  \label{eq:abeldirect}
\end{equation} 
$x=R/b$,  and $a=b\,\rho(0)$. The variable $a$ has units of surface density. In general,  for an arbitrary index $m$, the function $f(x,m)$ has to be evaluated numerically. However, $f$ has a close analytic expression for  $m=5$ \citep{2021ApJ...921..125S}, namely,
\begin{equation}
  f(x,5) = \frac{4/3}{(1+x^2)^2},
  \label{eq:ppoly5}
\end{equation}
which has its half-mass radius at $x=1$.  As it happens with the polytropes, all \propol s show a central plateau \citep[see][Fig.~1, as well as Fig.~\ref{fig:lane_emden_fitsersic} in here]{2021ApJ...921..125S}.
%
%
%

\section{Relation between polytropes and \sersic\ profiles}\label{sec:sersic}
The observed stellar mass surface density within galaxies drops with radial distance following a law  approximately given by \sersic\ functions \citep{1968adga.book.....S,2005PASA...22..118G},
\begin{equation}
  \Sigma(R)= \Sigma(0)\,\exp{\large[}-c_n\,(R/R_e)^{1/n}{\large]},
  \label{eq:sersicdef}
\end{equation}
with $R_e$ the radius enclosing half of the mass and $c_n$ a constant which depends only on the so-called \sersic\ index $n$. The \sersic\ index controls the shape of the profile, and has been observed to vary from 0.5 to 6 \citep{2003ApJ...594..186B,2012ApJS..203...24V}, approximately going from disk-like galaxies  \citep[$n\simeq 1$;][]{1994A&AS..106..451D} to elliptical galaxies  \citep[$n\simeq 4$;][]{1948AnAp...11..247D}.
The question arises as to whether the theoretical polytropes account for the empirical \sersic\ profiles. Since \sersic\ functions describe surface densities they have to be compared with \propol s (Sect.~\ref{sec:propols}). Both \propol s and \sersic\ profiles present cores, however, they are not similar. Except for $n\lesssim 1$, the cores of the \sersic\ profiles are too small compared with the cores of the \propol s \citep{2021ApJ...921..125S}. This is shown in Fig.~\ref{fig:lane_emden_fitsersic}, which shows three different Sersic profiles covering the whole range of observed indexes together with fits to them using \propol s. There is a large mismatch in the cores, but their outskirts are indistinguishable within any realistic observational error.  
The equivalence between \sersic\ index $n$ and polytropic index $m$ depends somewhat on the range of radii used for comparison, but it approximately goes from   $m\simeq 2$ for  $n=0.5$ to  $m\simeq 5$ for $n=6$ \citep{2021ApJ...921..125S}. Thus, the range of physically sensible polytropes (Eq.~[\ref{eq:nlimits}]) seem to naturally yield the range of observed \sersic\ indexes  \citep{2003ApJ...594..186B,2012ApJS..203...24V}.  Because of the mismatch in the cores, it is at present unclear whether the range of observed \sersic\ indexes truly arises from the range of sensible polytropic indexes. 

\sersic\ profiles describe stellar mass whereas \propol s model total mass, i.e., gas, stars, and DM all together. For the similarity between \sersic\ profiles and \propol s to be  of relevance for real galaxies, a scaling between stellar mass and total mass must exist. Fortunately, the similarity still holds even when the scaling is not a constant and the ratio between stellar and total mass varies radially \citep[see][]{2021ApJ...921..125S}.

%
\section{Relation between polytropes and CDM mass density profiles}\label{sec:nfwproperties}
According to the current cosmological model, the DM provides most of the gravitational force that drives galaxy formation and evolution. The DM particles in this model are {\em cold}, and only interact with each other and with the baryons through gravity \citep[for an up-to-date critical review, see][]{2021arXiv210602672P}. Given the expected low mass of the DM particles \citep{2020ARNPS..70..355C}, gravitation alone cannot thermalize the particle distribution within a Hubble time \citep{2008gady.book.....B} so the current DM mass distribution in galaxy haloes should still reflect the initial conditions.
Back in \citeyear{1997ApJ...490..493N},  \citet{1997ApJ...490..493N} found that the DM haloes in numerical simulations follow a universal law, now called NFW profile, where the density drops with radius $r$ as
\begin{equation}
  \rho_{\rm NFW}(r)=\frac{4\,\rho_{\rm NFW}(r_s)}{(r/r_s)(1+r/r_s)^2},
  \label{eq:nfw}
\end{equation}
where $r_s$ is a characteristic radius. Because the NFW profile results from cold DM numerical simulations, its shape is commonly explained as the outcome of the cosmological initial conditions \citep{2004MNRAS.352.1109A,2014ApJ...790L..24C,2015ApJ...805L..16N,2017MNRAS.465L..84L,2020MNRAS.495.4994B}.
 It is repeatedly found in literature that DM halos are reproduced by Einasto profiles as well as, if not better than, the NFW profiles \citep{2004MNRAS.349.1039N,2005ApJ...624L..85M,2013MNRAS.428.2805A}. The Einasto profiles, $\rho_{ei}$, are formally identical to a \sersic\ profile but with the variable representing the 3D radial distance, $r$, rather than the projected distance, $R$, i.e.,
\begin{equation}
  \rho_{ei}(r) = B\,\exp(-A\,r^\mu),
  \label{eq:einasto}
\end{equation}
where $A$, $B$, and $\mu$ are the three parameters that define the profile. 
\begin{figure}
  \centering 
\includegraphics[width=0.65\linewidth]{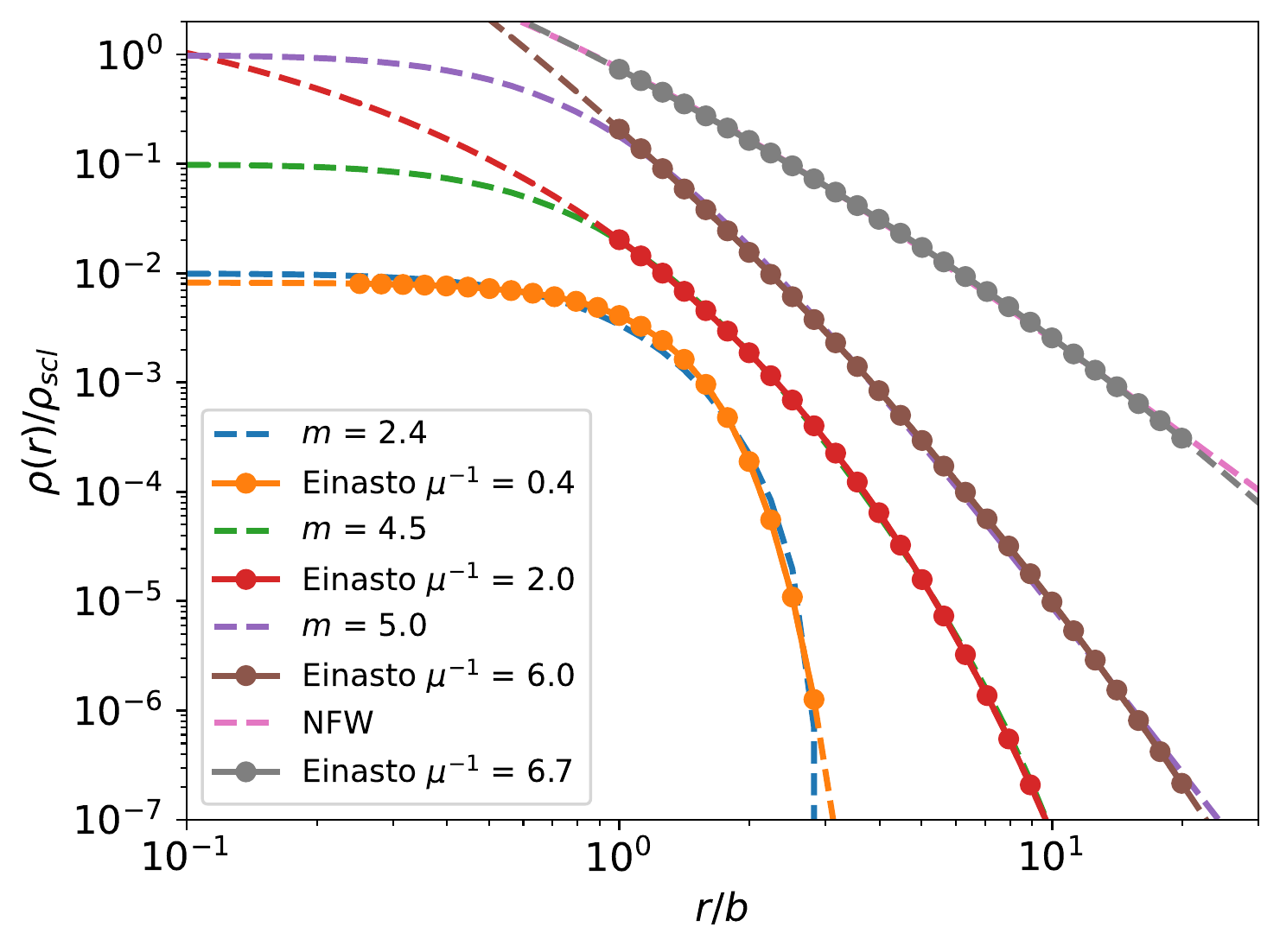}
\caption{Representative polytropes  (dashed lines, as indicated in the inset) fitted with Einasto functions (dashed lines and symbols of  the same color).  The symbols represent the range of radii included in the fit. Outside the core, the agreement is well within any realistic observational error, an agreement holding for up to a factor $\sim\,20$ in radius and $\sim\,10^5$ in surface density.  The  original polytropic index and the corresponding inverse exponent of the Einasto function ($\mu^{-1}$) are given in the inset. For display purposes, all profiles have different normalization ($\rho_{scl}$) and are scaled radially to the scale-length of the corresponding polytrope ($b$). The NFW profile (the pink dashed line) is well fitted by an Einasto profile (grey dots and dashed line), but the shape of this Einasto profile differs from any polytrope of finite size and mass.  
}
\label{fig:lane_emden_fiteinasto}
\end{figure}
Figure~\ref{fig:lane_emden} includes a NFW profile (the orange dashed line) with $r_s$ and $\rho_{\rm NFW}(r_s)$ set arbitrarily to fit in within the panel. The NFW profile diverges towards the center of the mass distribution, $\rho_{\rm NFW}\rightarrow\infty$ when $r\longrightarrow 0$, and in this sense is very different from a polytrope where the density has a central plateau (Eq.~[\ref{eq:central2}]) . The same statement holds for the Einasto profiles. Their cores are not like a polytropic core unless $\mu = 2$\footnote{Compare the expansion of Eq.~(\ref{eq:einasto}) when $r\rightarrow 0$ with Eq.~(\ref{eq:central2}).}. However, as it happens with the \sersic\ profiles (Sect.~\ref{sec:sersic}) when they are compared with \propol s, the outskirts of the Einasto profiles are very close to a polytrope (Fig.~\ref{fig:lane_emden_fiteinasto}). This is not the case of the NFW profiles, which do not seem to have a correspondence in the realm of polytropes. This seems to be in contradiction with the claim that both NFW and Einasto profiles reproduce DM haloes from numerical simulations, but it is not. The Einasto profile equivalent to a NFW in simulations has a particular $\mu\simeq 0.15$ \citep[e.g.,][see also Fig.~\ref{fig:lane_emden_fiteinasto}]{2004MNRAS.349.1039N,2020Natur.585...39W}, which would correspond to a polytrope with $m \sim 7$, i.e., far from the polytropes with finite size and mass (see Fig.~\ref{fig:lane_emden}).
Cosmological numerical simulations of collision-less DM particles produce halos without core (Eq.~[\ref{eq:nfw}]). However, when collisions are included, then the resulting DM haloes always develop a polytropic central core \citep{2021MNRAS.504.2832S}. There are two ways in which collisions between particles have been considered in simulations. One of them is actually an artifact due to the need of using DM particles with artificially large masses. Then the two-body gravitationally induced collisions become unrealistically important, and the inner DM halo has to be discarded to recover proper NFW profiles \citep{2003MNRAS.338...14P}. On the other hand, collisions are imposed to model the so-called  self-interacting DM \citep{2000PhRvL..84.3760S}, finding that the shape of the resulting halo has a core whose shape is independent of the assumed collisional cross-section \citep[][]{2015MNRAS.453...29E}. In both cases the resulting DM haloes have a polytropic core (see Fig.~\ref{fig:polycore}). This result reinforces the ansatz that the Tsallis entropy is indeed adequate to describe self-gravitating systems in thermal equilibrium.    
\begin{figure}
  \centering 
 \includegraphics[width=0.5\linewidth]{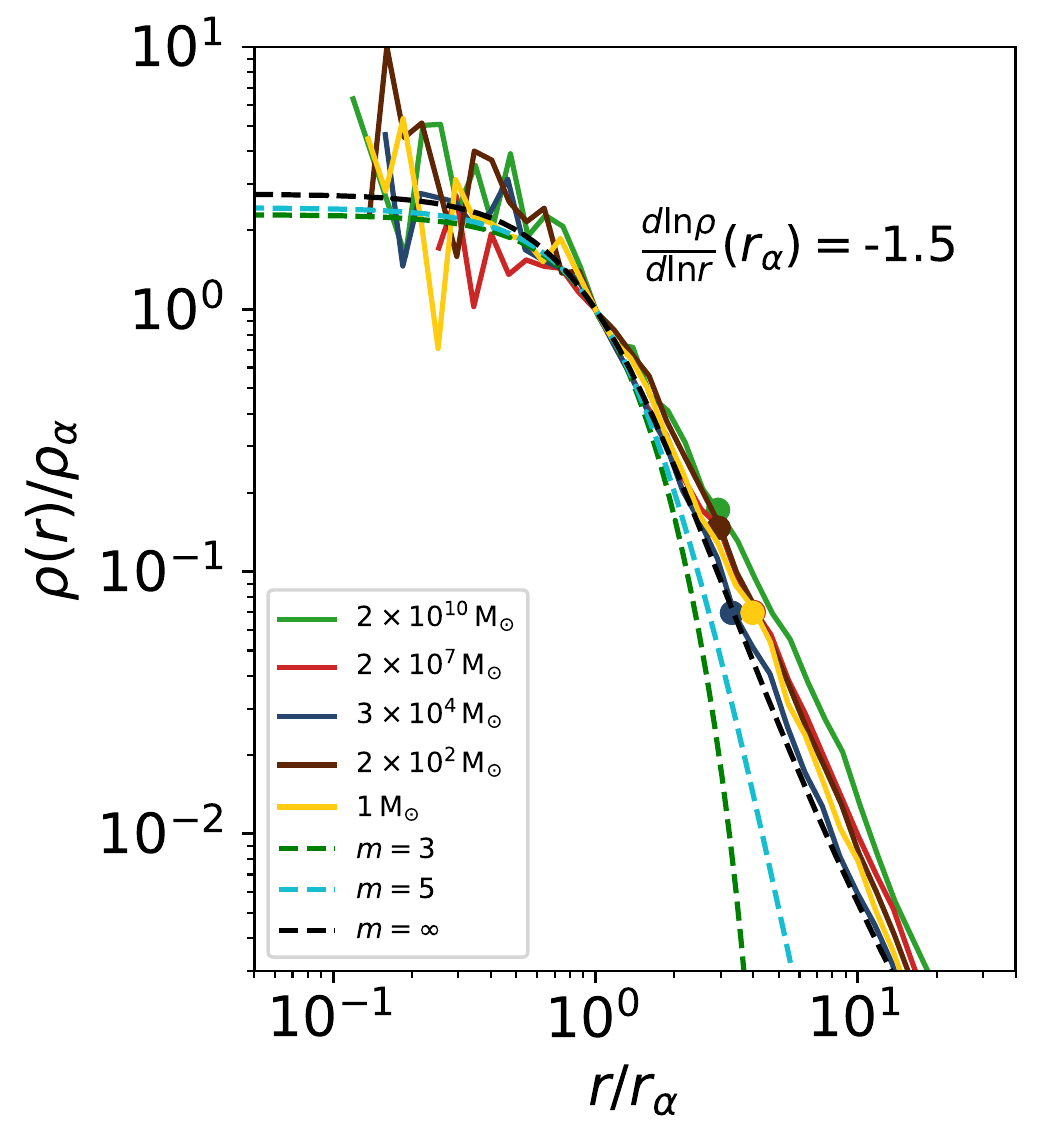}
\caption{Mass density profiles of DM halos from numerical simulations showing the artificial cores appearing within the convergence radius (marked by color symbols). Within this radius, the timescale for two-body collisions between the (artificially massive) DM particles used in the simulation is shorter than the age of the Universe. The profiles have been taken from \citep{2020Natur.585...39W}  and they represent DM haloes with masses differing by as much as 10 orders of magnitude (see the inset). The dashed lines represent
  polytropes, with their index included in the inset. All profiles have been normalized to the density and radius where the logarithmic derivative of the profile (Eq.~[\ref{eq:logder}]) equals -1.5. Additional details can be found in \citep{2021MNRAS.504.2832S}.}
    \label{fig:polycore}
  \end{figure}

Claims that polytropes reproduce simulated DM halos better than NFW profiles can be found in literature \citep{2006AIPC..857..316N}.  

%
%
  \section{Observational support for galaxies reaching Tsallis maximum entropy}\label{sec:obs_support}

  %
  %

The use of Tsallis entropy is not new to the literature on galaxies, and it has already shown quite some success. It provides a common framework linking a number of seemingly disconnected observational properties of galaxies. 
The mass density profiles observed in the centers of dwarf galaxies are very well reproduced by polytropes without any degree of freedom or tuning \citep{2020A&A...642L..14S}. In other words, the so-called {\em core-cusp} problem of the  CDM cosmology (see Sect.~\ref{sec:intro}) is automatically solved if the mass distribution within dwarfs is in TE as defined by the Tsallis entropy. Similar conclusions have also being found elsewhere \citep{2021A&A...647A..29N}.
The same type of profile also explains the stellar surface density profiles observed in globular clusters  \citep{trujillo21}.

Claims in the literature that polytropes do not provide good fits to observed galaxies \citep[e.g.,][]{2005PhyA..356..419C,2008PhRvE..77b2106F} have been disproved lately \citep[e.g.,][]{2021A&A...647A..29N,2021ApJ...921..125S}. It has been shown that \propol s account for the mass distribution in low mass galaxies ($M_\star < 10^9\,{\rm M}_\odot$), and they do it better than classical \sersic\ profiles \citep{2021ApJ...921..125S}.  The goodness of \propol s to reproduce the observed mass distribution in galaxies parallels the equivalence between \propol s and \sersic\ profiles analyzed in Sect.~\ref{sec:sersic}.  Although with scatter, the observed polytropic indexes increase with increasing mass and tend to cluster around $m = 5$. For the most massive galaxies, \propol s are very good at reproducing their central parts, but they do not handle well cores and outskirts overall. It seems like if the central parts are already in TE, a condition not reached in the outskirts yet \citep[][]{2004MNRAS.355.1217H}.

The range of physically sensible polytropes (Eq.~[\ref{eq:nlimits}]) seems to naturally yield the range of observed \sersic\ indexes  \citep[from 0.5 to 6;][]{2003ApJ...594..186B,2012ApJS..203...24V}. As we discuss in Sect.~\ref{sec:nfwproperties},  because of the mismatch in the cores between \propol s and  \sersic s, it is at present unclear whether this agreement is pure coincidence or if it provides the long-sought explanation for why observed galaxies follow \sersic\ profiles. Even with this caveat, we argue that the equivalence between sensible polytropic indexes and observed \sersic\ indexes supports that galaxies develop polytropic mass distributions.

%

The star counts and the kinematic data of the Milky Way (MW) stellar halo are well represented by an Einasto profile with $\mu^{-1} \simeq 2$ and an effective radius $\simeq 20$\,kpc \citep{2014MNRAS.443..791E}. Because of the equivalence between Einasto profiles and polytropes (Sect.~\ref{sec:nfwproperties} and Fig.~\ref{fig:lane_emden_fiteinasto}), this MW halo corresponds to a polytrope with $m$ between 4 and 5.


%
%
%
  \section{Pathways to thermalization}\label{sec:thermalization}
  
As we put forward in Sect.~\ref{sec:obs_support}, some observations suggest that TE, as described by the Tsallis entropy, sets the internal distribution of mass in some galaxies. Therefore, the arguments in Sect.~\ref{sec:intro} against galaxies being able to reach TE are questionable. In particular, there should be alternatives to the two-body relaxation collisions, a mechanism unable to thermalize the gravitational potential within the Hubble time. What are these alternative pathways to thermalization?

{\em Violent relaxation} \citep{1967MNRAS.136..101L,2008gady.book.....B} can do the work. If for some reason the self-gravitating system is driven far from equilibrium, then the gravitational potential varies in time, and this variation allows the particles to exchange their energy and momentum in a timescale comparable with the dynamical timescale for the variation of the potential. For instance, if the system starts off out the virial equilibrium and collapses, this timescale will be set by the free-fall time, which is much shorter than the Hubble time\footnote{For a MW-like halo of mass and radius $10^{13} {\rm M}_\odot$ and 100\,kpc, respectively, the free-fall time is only 0.2~Gyr.}. This idea permeates various physical processes invoked to turn the {\em cusps} expected from the CDM simulations into the observed {\em cores}. For example, the feedback of the baryons on the DM particles through gravitational forces \citep[][]{2010Natur.463..203G,2014MNRAS.437..415D,2020MNRAS.499.2912F}. Supernova explosions suffice to expel a significant fraction of the gas existing in the central regions of dwarf galaxies in a very short timescale, changing the overall gravitational potential and forcing the DM distribution to readjust. Gas gets re-accreted, new stars are formed, and the process starts over rendering a cored density profile after several of these cycles.

Another pathway to thermalization may be through the mergers of SMBHs expected to occur at the center of massive ellipticals \citep[][]{2014ApJ...795L..31L,2016MNRAS.462.2847M}. The motion of two merging black holes produces scouring of stars. In addition, the recoil kicks the merged SMBH out of the center, forcing a final swing of the SMBH that stirs the global gravitational potential \citep{2006RPPh...69.2513M,2021MNRAS.502.4794N}.  Thus, scouring plus recoil may allow the self-gravitating system to reach TE in a timescale much shorter than the two-body relaxation timescale. The fast scouring of the inner region can also be achieved through scattering with massive gas clumps \citep[][]{2013ApJ...775L..35E,2019MNRAS.489.5919S} or by the forcing produced by a central bar \citep{1971ApJ...168..343H}. We note that central BHs may also be present in dwarf galaxies \citep[e.g.,][]{2017IJMPD..2630021M,2022arXiv220109903D}, therefore, this thermalizing process may work for them as well.

Another extremely interesting pathway to thermalization has to do with the unknown nature of the DM particles. Thermalization in a short timescale is possible if two-body collisions between particles are efficient enough. This would happen if DM particles had masses in the stellar-mass realm (e.g., if they are stellar-mass primordial BHs), a possibility seemingly discarded by observations \citep[][]{2020ARNPS..70..355C}. A more appealing  possibility is DM not being collision-less, i.e., having an additional large DM particle -- DM particle collision cross section that shortens the two-body collision timescale below the Hubble time \citep{2000PhRvL..84.3760S,2001ApJ...547..574D,2015MNRAS.453...29E}.
When the DM particles of numerical simulations are allowed to interact through any of these two mechanisms, it leads to a gravitational potential conforming with polytropes \citep[][]{2021MNRAS.504.2832S}. 

%

%
\section{Conclusions}\label{sec:conclusions}

We still do not have a final answer to the original question of  {\em  why do real galaxies choose a particular mass distribution?}
Here we have examined whether {\em  thermodynamic equilibrium (TE) is setting the mass distribution}.
The standard answer discards the role of  TE for a number of reasons.
Firstly, the TE described by the classical Boltzmann-Gibbs entropy produces mass distributions with unphysical properties (Sect.~\ref{sec:BGentropy}).
Secondly, two-body gravitational collisions are quite inefficient and so TE cannot be set and should not be relevant (Sect.~\ref{sec:intro}).
Finally, the DM halos coming from numerical simulations (i.e., the NFW profile and its relatives) have a mass distribution set by initial conditions rather than by TE  (Sect.~\ref{sec:intro}). 
However, these arguments are questionable for a number of reasons.
Firstly, the  Boltzmann-Gibbs entropy does not describe systems with long range interaction. When the Tsallis entropy is used to define TE, then the resulting mass distribution (polytropes; Sect.~\ref{sec:tsallis}) turns out to be physically sensible.  
Secondly, there are alternatives to the two-body collisions to thermalize the potential. They go from feedback of baryons on DM through gravity to the stirring of the potential produced by the merging of SMBHs (Sect.~\ref{sec:thermalization}).
Thirdly, the standard DM halos produced in numerical simulations assume collision-less DM particles thus, by construction, they have to reflect initial conditions rather than TE. However, when collisions are allowed, the numerical haloes are consistent with polytropes and so with TE (Sect.~\ref{sec:nfwproperties}). This result reinforces the ansatz that the Tsallis entropy is indeed adequate to describe self-gravitating systems in thermal equilibrium.

In addition to the above arguments rebutting the original criticisms,  a number of observations indicate that polytropic profiles reproduce real galaxies (Sect.~\ref{sec:obs_support}). In particular, the outskirts of plane-of-the-sky projected politropes are extremely similar to \sersic\ profiles (Sect.~\ref{sec:sersic}), and they do reproduce the stellar mass distribution in galaxies. Even the halo of the MW seems to be a polytrope.  Moreover, the cores in the observed mass distribution of dwarf galaxies are polytropes so solving the so-called  {\em core -- cusp} problem seems to be equivalent to explaining why the mass distribution in these objects is thermalized (Sects.~\ref{sec:intro} and \ref{sec:obs_support}).   

Section~\ref{sec:tsallis} summarizes a number of properties of polytropes that are of interest in the context of galaxy structure. In particular, the cores of all polytropes have the same shape, independently of polytropic index $m$ (Eq.~[\ref{eq:central2}]), the velocity dispersion scales as a power of the density  (Eq.~[\ref{eq:contrast}]), and  there is a fairly narrow range of valid values of $m$ (Eq.~[\ref{eq:nlimits}]).

As a way to foretell future developments in the field, I conclude with a few interesting and open lines of research. Ordered from more general to more specific, they are: 
(1) From first principles, derive the  entropy describing self-gravitating systems \citep[][]{2011Entrp..13.1765T,2018Entrp..20..813A}.
(2) Include angular momentum in the polytropic formalism \citep[][]{1986ApJ...300..112B,2013MNRAS.436.2014N,2017MNRAS.467.5022H}.
(3) Are the haloes in fully DM dominated galaxies thermalized? I so, then the DM cannot be collision-less as assumed in the standard cosmological model \citep[][]{2014MNRAS.437..415D,2015MNRAS.454.2092O,2021arXiv210602672P}.
(5) Observationally, the central mass surface density in galaxies seems to have an upper limit \citep[$\Sigma(0) \lesssim 75\,{\rm M_\odot\,pc^{-2}}$, e.g.][]{2000ApJ...537L...9S,2020ApJ...904..161B}. Is this a consequence of the galaxies being polytropes\citep{2002A&A...386..732C}?

\vspace{6pt} 




\funding{
This research was partly funded by the Spanish Ministry of Science and Innovation, project  PID2019-107408GB-C43 (ESTALLIDOS), and by Gobierno de Canarias through EU FEDER funding, project PID2020010050.
}

\acknowledgments{
Most of what is described in these notes follows from conversations with various colleagues: 
Ignacio Trujillo,  Angel R. Plastino,  Ana Monreal Ibero, Claudio Dalla Vecchia, Diego Blas, and Jorge Mart\'\i n Camalich.
Aridane Rodr\'\i guez Moreno helped me to the compile the references cited in {\em Analytic approximations}, Sect.~\ref{properties_pol}.
}

\conflictsofinterest{The author declares no conflict of interest.}



\abbreviations{Abbreviations}{
  The following abbreviations are used in this manuscript:\\
  
\noindent 
\begin{tabular}{@{}ll}
BH & Black Hole\\
CDM & Cold Dark Matter\\
DF & Distribution Function\\
  DM & Dark Matter\\
  MW & Milky Way\\
  NFW & Navarro, Frenk, and White \citep{1997ApJ...490..493N}\\
  SMBH & Super Massive Black Hole\\
  TE & Thermodynamic Equilibrium
\end{tabular}
}

\reftitle{References}


\newcommand{\prd}{Physical Review D}
\newcommand{\pre}{Physical Review E}
\newcommand{\prl}{Phys. Rev. Lett.}
\newcommand{\apjl}{ApJL}
\newcommand{\aj}{AJ}
\newcommand{\aap}{A\&A}
\newcommand{\aaps}{A\&AS}
\newcommand{\apj}{ApJ}
\newcommand\apjs{ApJS}
\newcommand{\mnras}{MNRAS}
\newcommand{\pasa}{PASA}
\newcommand{\jcap}{JCAP}
\newcommand{\nat}{Nat}
\newcommand{\ssr}{SSRv}
\newcommand{\araa}{ARA\&A}
\newcommand\apss{Ap\&SS}
\externalbibliography{yes}
\bibliography{universe_references}


%


\end{document}